\documentclass[12pt]{article}

\usepackage{graphicx,amssymb,url,mathrsfs, amsmath}
\usepackage{eucal}
\usepackage{wrapfig}
\usepackage{boxedminipage}
\usepackage{setspace}
\usepackage{subfigure}
\def\a  {\alpha}                
       \def\d  {\delta}        \def\D  {\Delta}
\def\e  {\epsilon}        \def\k  {\kappa}
\def\l  {\lambda}             \def\m  {\mu}
\def\n  {\nu}          \def\s  {\sigma}        
\def\t  {\tau}                 
   \def\w  {\omega}

\newcommand{\cala}{\mbox{${\cal A}$}} 
 
 \newcommand{\calf}{\mbox{${\cal F}$}}

\def\IR{{\hbox{{\rm I}\kern-.2em\hbox{\rm R}}}}
\def\IB{{\hbox{{\rm I}\kern-.2em\hbox{\rm B}}}}
\def\IN{{\hbox{{\rm I}\kern-.2em\hbox{\rm N}}}}
\def\IC{\,\,{\hbox{{\rm I}\kern-.59em\hbox{\bf C}}}}
\def\IZ{{\hbox{{\rm Z}\kern-.4em\hbox{\rm Z}}}}
\def\IP{{\hbox{{\rm I}\kern-.2em\hbox{\rm P}}}}
\def\IH{{\hbox{{\rm I}\kern-.4em\hbox{\rm H}}}}
\def\ID{{\hbox{{\rm I}\kern-.2em\hbox{\rm D}}}}

\def\be{\begin{equation}}
\def\ee{\end{equation}}
\def\ba{\begin{eqnarray}}
\def\ea{\end{eqnarray}}

\def\half{\frac{1}{2}}

\newcommand{\inv}[1]{\frac{1}{#1}}

\newcommand{\ud}{\mbox{${\mathrm{d}}$}}

\def\dell{\partial}

\newcommand{\abs}[1]{\left| #1 \right|}

\def\dg{{\dagger}}
\def\Tr{{\rm tr}\,}

\def\nn{\nonumber}
\def\ea{{\it et al}. }

\newcommand{\Mkk}{M_{\rm KK}}
\newcommand{\wt}{\widetilde}
\newcommand{\wh}{\widehat}

\begin{document}

\begin{titlepage}

%\begin{flushright}
%  {\tt hep-th/0608046}
%\tt {FileName:FF1.tex} \\
% {\tt \today}
%\end{flushright}
\vspace{0.5in}

\begin{center}
{\large \bf Holographic Nucleons~\footnote{To appear in {\it Festschrift for Gerry Brown}, 
Ed. Sabine Lee, World Scientific}} \\
\vspace{10mm}
Ismail Zahed\\
\vspace{5mm}
{\it  Department of Physics and Astronomy, SUNY Stony-Brook, NY 11794}\\
\vspace{10mm}
  {\tt \today}
\end{center}
\begin{abstract}
Recent developments in holography have provided a new vista to the nucleon 
composition. A strongly coupled core nucleon tied with vector mesons emerge 
in line with the Cheshire cat principle. The cat is found to hide in the holographic
direction. We discuss the one, two and many baryon problem in this context and
point at the striking similarities between the holographic results and recent lattice
simulations at strong coupling.
\end{abstract}

\end{titlepage}

\renewcommand{\thefootnote}{\arabic{footnote}}
\setcounter{footnote}{0}

%\tableofcontents
%\newpage

%%%%%%%%%%%%%%%%%%%%%%%%%%%%%%%%%%%%%%%%%%%%%%%%%%%%%%%%%%%%%%%%%%%%%%%%%%%
%%%%%%%%%%%%%%%%%%%%%%%%%%%%%%%%%%%%%%%%%%%%%%%%%%%%%%%%%%%%%%%%%%%%%%%%%%%
%%%%%%%%%%%%%%%                 BODY                    %%%%%%%%%%%%%%%%%%%
%%%%%%%%%%%%%%%%%%%%%%%%%%%%%%%%%%%%%%%%%%%%%%%%%%%%%%%%%%%%%%%%%%%%%%%%%%%
%%%%%%%%%%%%%%%%%%%%%%%%%%%%%%%%%%%%%%%%%%%%%%%%%%%%%%%%%%%%%%%%%%%%%%%%%%%

\section{Dedication}
{\it This paper is dedicated to Gerry Brown who has been my mentor
and colleague for the past three decades. I have met Gerry while in
graduate school way back at MIT, in Feshbach's office on a sunny
fall morning. After the meeting, Gerry asked me when I would graduate
as he was prepared to hire me for the next three years. I did the following spring
and have been  with Gerry  since then. Gerry is an outstanding scientist and humanist 
that has contributed immensely to our field. I thank him for his 
guidance and friendship, and wish him well  for the years to come.}

\section{Introduction}

Back in the eighties, quark bag models were proposed as models
for hadrons that capture the essentials of asymptotic freedom
through weakly interacting quarks and gluons within a bag, and
the tenets of nuclear physics through strongly interacting 
mesons at the boundary. The delineation or bag radius was 
considered as a fundamental and physically measurable scale
that separates ultraviolet from infrared QCD. Two competing
pictures emerged: The original MIT bag model with a large
radius surrounded by a bare vacuum and the Stony-Brook bag
model with a small radius surrounded by pions~\cite{BAGS}.  
In fact, at low energy this delineation is unphysical as stated
in the Cheshire cat principle~\cite{CAT}. Quantum effects and
anomalies cause most of the charges (fermionic, axial, etc.)
to leak making the bag boundary immaterial~\cite{LEAK}, much 
like the smile of the Cheshire cat in "Alice in wonderland"~\cite{ALICE}.
The Skyrme model typifies the extreme realization of the Cheshire cat
principle whereby the immaterial bag radius is reduced to zero size
\cite{SKYRMION}.

The Skyrme model realizes QCD baryons as chiral solitons 
in the limit of large number of colors $N_c$.  Recently, the same 
model was found to emerge from holographic QCD in the dual limit
of large $N_c$ and strong coupling t'Hooft coupling $\l=g^2N_c$
~\cite{SAKAI,HRYY}. In the holographic construction, the Skyrmion 
is the holonomy of a flavor instanton tied to Witten's vertex 
in bulk~\cite{WITTEN}. Baryon number at the boundary is dual to instanton 
number in bulk. Although the exact holographic dualities are only known 
for a restricted set of string theories in bulk with mostly
conformal field theories at the boundary~\cite{MALDACENA}, we will 
assume that such a correspondence holds for 
holographic QCD which is not conformal.

The present paper review some aspects of the holographic baryons
following recent work in~\cite{KZ}. It is dedicated to Gerry Brown.
In section 3, we review the holographic baryon construction from a bulk
instanton, and emphasize the emergence of the Cheshire cat principle.
In section 4 a top-down holographic model is briefly summarized
and the bayonic current derived. In section 5, the 2-nucleon problem 
is discussed usingthe ADHM 2-instanton configuration.  In section 6, 
cold and dense holographic matter is argued to be a crystal of instantons
at low densities, and a crystal of dyons at higher densities. In section 7,
we estimate the melting temperature of these crystals into liquids. 
Our conclusions are summarized in section 8.

\begin{figure}[]
  \begin{center}
    \includegraphics[width=11cm]{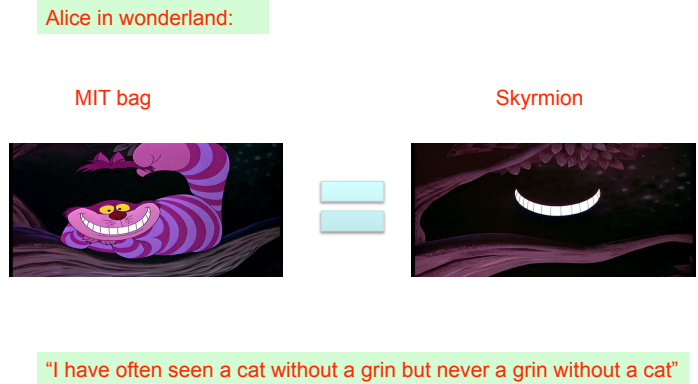}
  \caption{Alice's Cheshire Cat.}
  \label{Fig:fig1XX}
  \end{center}
\end{figure}
\begin{figure}[]
  \begin{center}
    \includegraphics[width=11cm]{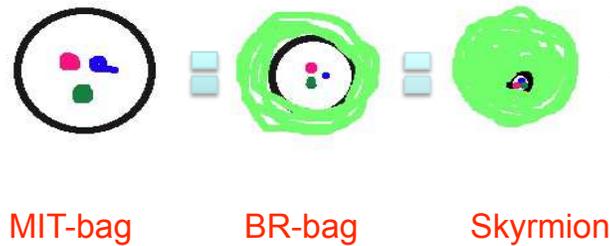}
  \caption{The Cheshire Cat Principle.}
  \label{Fig:fig1X}
  \end{center}
\end{figure}

\section{The Skyrmion from the Instanton}

In holographic QCD, a baryon is initially described as a flavor
instanton in an $R^{1,3}\times R_Z$ dimensional space where
the holographic direction $R_Z$ is warped by gravity. At large
t'Hooft coupling $\l=g^2N_c$, the instanton is forced by gravity to 
drop to the bottom of $R_Z$. Topological (Coulomb) repulsion
causes the instanton to lump at a size $Z\approx 1/\sqrt{\l}\ll 1$. 
In this limit, the effects of the gravitational warping on the instanton
can be neglected. Thus the instanton SU(2) flavor configuration
${\mathbb{A}}_M$ and its supporting U(1) Coulomb potential
$\widehat{\mathbb{A}}_M$ read

\begin{eqnarray}
\widehat{\mathbb{A}}_0 = -\frac{1}{8 \pi^2 a \l}\frac{2\rho^2 +
\xi^2}{(\rho^2 + \xi^2)^2} \ , \qquad\qquad\qquad
\mathbb{A}_M = \eta_{iMN}\frac{\sigma_i}{2}\frac{2
x_N}{\xi^2 + \rho^2} \ ,   
\label{ADHM2}
\end{eqnarray}
with all other gauge components zero. The size is $\rho\sim 1/\sqrt{\l}$.
We refer to~\cite{SAKAI} (last reference) for more details on the relevance 
of this configuration for baryons. The ADHM configuration has maximal spherical 
symmetry and satisfies
\begin{eqnarray}
(\mathbb{R}\mathbb{A})_Z = \mathbb{A}_Z(\mathbb{R} \vec{x}) \ ,
\qquad (\mathbb{R}^{ab}\mathbb{A}^{b})_i =
\mathbb{R}^T_{ij}\mathbb{A}_j^a(\mathbb{R}\vec{x})\,\,,
\end{eqnarray}
with $\mathbb{R}^{ab}\tau^b=\Lambda^+\tau^a\Lambda$ a rigid SO(3) rotation,
and $\Lambda$ is SU(2) analogue..

The holographic baryon is just the holonomy of (\ref{ADHM2}) along the
gravity bearing and conformal direction $Z$,

\begin{eqnarray}
U^{\mathbb{R}}(x)=\Lambda{\bf P}{\rm exp}
\left(-i\int_{-\infty}^{+\infty}dZ\,\mathbb{A}_Z\right)\Lambda^+\,\,.
\label{HOLO}
\end{eqnarray}
The corresponding Skyrmion in large $N_c$ and leading order
in the strong coupling $\l$ is
$U(\vec{x})=e^{i\vec{\tau}\cdot\vec{x}{\bf F} (\vec{x})}$ with
the profile

\begin{eqnarray}
{\bf F} (\vec{x})=\frac{\pi |\vec{x}|}{\sqrt{{\vec{x}}^2+\rho^2}}\,\,.
\label{FF}
\end{eqnarray}
The holonomy (\ref{HOLO}) is a heavy flavor but colorless 
fermion "propagating" along the holographic direction with 
the instanton as a background field. The result is a small size
Skyrmion map $U(x)$ at the boundary.

The baryon emerges from a semiclassical organization of the quantum fluctuations
around the point-like source (\ref{HOLO}). To achieve this, we define

\begin{eqnarray}
A_M(t,x,Z) = \mathbb{R}(t) \left(\mathbb{A}_M(x-X_0(t),Z-Z_0(t))
+ C_M(t,x-X_0(t),Z-Z_0(t))\right) \ , \label{SEM1}
\end{eqnarray}
The collective coordinates 
$\mathbb{R}, X_0, Z_0, \rho$ and the fluctuations $C$ in 
(\ref{SEM1}) form a redundant set. The redundancy is lifted 
by constraining the fluctuations to be orthogonal to the
zero modes. This can be achieved either rigidly~\cite{ADAMI} 
or non-rigidly~\cite{VERSHELDE}. We choose the latter as it 
is causality friendly. For the collective iso-rotations the
non-rigid constraint reads
\begin{eqnarray}
\int_{x=Z=0} d{\hat{\xi}} C\,G^B\mathbb{A}_M\,\,,
\label{SEM5}
\end{eqnarray}
with $(G^B)^{ab}=\epsilon^{aBb}$ the real generators of $\mathbb{R}$.

\begin{figure}[]
  \begin{center}
    \includegraphics[width=11cm]{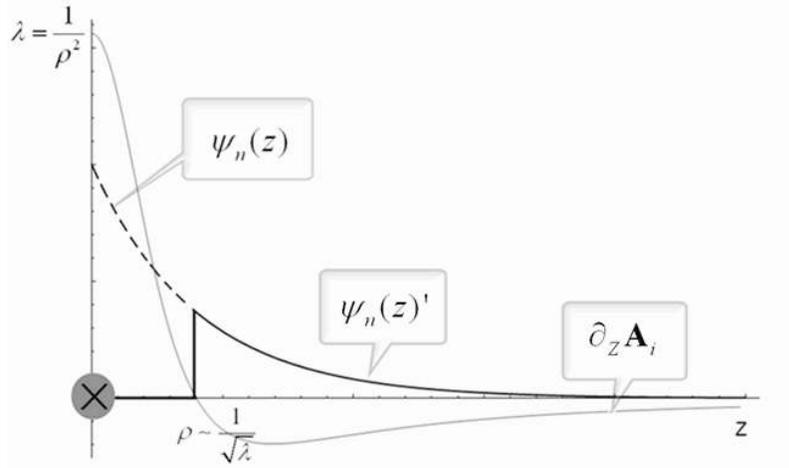}
  \caption{The Z-mode in the non-rigid gauge vs $\dell_Z \mathbb{A}_i$.}
  \label{Fig:fig1}
  \end{center}
\end{figure}

For $Z$ and $\rho$ the non-rigid constraints are more natural to
implement since these modes are only soft near the origin at large
$\l$. The vector fluctuations at the origin linearize through the modes
\begin{eqnarray}
d^2\psi_n/dZ^2= -\l_n\psi_n \ , \label{SEM6}
\end{eqnarray}
with $\psi_n(Z)\sim e^{-i\sqrt{\l_n}Z}$. In the spin-isospin 1
channel they are easily confused with $\partial_Z\mathbb{A}_i$
near the origin as we show in Fig.~\ref{Fig:fig1}.
Using the non-rigid constraint,  the double counting is removed
by removing the origin from the vector mode functions
\begin{eqnarray}
\psi_n'(Z)=\theta(|Z|-Z_C)\psi_n(Z) \ , \label{SEM7}
\end{eqnarray}
with $Z_C\sim\rho\sim 1/\sqrt{\l}$ which becomes the origin for
large $\l$. In the non-rigid semiclassical framework, the baryon
at small $\xi<|Z_C|$ is described by a flat or uncurved
instanton located at the origin of $R^4$ and rattling in 
the vicinity of $Z_C$. At large $\xi>|Z_C|$, the rattling instanton 
sources the vector meson
fields described by a semi-classical expansion with non-rigid
Dirac constraints. Changes in $Z_C$ (the core boundary) are
reabsorbed by a residual gauge transformation on the core
instanton. This is a holographic realization of the Cheshire cat
principle~\cite{CAT} where $Z_C$ plays the role of the Cheshire
cat smile. In a way, Alice's Cheshire cat of Fig.~\ref{Fig:fig1XX} has 
gone out of sight in the holographic or 5th direction.

\section{The Baryonic Current}

To illustrate the Cheshire cat mechanism more quantitatively, we now
summarize the holographic Yang-Mills-Chern-Simons action in 5D 
curved background. This is the leading term
in a $1/\l$ expansion of the D-brane Born-Infeld (DBI)
action on D8~\cite{SAKAI},

\begin{eqnarray}
&&S = S_{YM} + S_{CS}\ ,  \label{YM-CS}\\
&&S_{YM} = - \k \int d^4x dZ \ \Tr \left[\half K^{-1/3}
\calf_{\m\n}^2 + \Mkk^2 K
\calf_{\m Z}^2 \right] \ , \label{YM} \\
&&S_{CS} = \frac{N_c}{24\pi^2}\int_{M^4 \times R}
\w_5^{U(N_f)}(\cala) \ , \label{CS}
\label{REDUCED}
\end{eqnarray}
where $\m,\n = 0,1,2,3$ are 4D indices and the fifth(internal)
coordinate $Z$ is dimensionless.  There are three things which are
inherited by the holographic dual gravity theory: $\Mkk, \k,$ and
$K$. $\Mkk$ is the Kaluza-Klein scale and we will set $\Mkk = 1$
as our unit. $\k$ and $K$ are defined as
\begin{eqnarray}
\k = {\l N_c} \inv {216 \pi^3} \equiv \l N_c a  \ , \qquad K = 1
+ Z^2 \ .
\end{eqnarray}
$\cala$ is the 5D $U(N_f)$ 1-form gauge field and $\calf_{\m\n}$
and $\calf_{\m Z} $ are the components of the 2-form field
strength $\calf = \ud \cala -i \cala \wedge \cala$.
$\w_5^{U(N_f)}(\cala)$ is the Chern-Simons 5-form for the $U(N_f)$
gauge field
\begin{eqnarray}
  \w_5^{U(N_f)}(\cala) = \Tr \left( \cala \calf^2 + \frac{i}{2} \cala^3 \calf - \inv{10} \cala^5
  \right)\ ,
\end{eqnarray}

We note that $S_{YM}$ is of order $\l$, while $S_{CS}$ is of order $\l^0$.
These terms are sufficient to carry a semiclassical expansion around
the holonomy (\ref{HOLO}) with $\hbar=1/\k$ as we now illustrate it for
the baryon current.

To extract the baryon current, we source the reduced 
action with $\hat{\cal V}_\mu(x)$ a $U(1)_V$ flavor 
field on the boundary in the presence of the vector
fluctuations ($C=\hat{v}$). The tree level baryonic 
current reads

\begin{eqnarray}
J^\m_{B}(x) &=& -  \k K
\widehat{\mathbb{F}}^{Z \m} (x,Z)\,
(1-\sum_{n=1}^\infty\a_{v^n}\psi_{2n-1})
\Big|_{Z=B}  \nn \\
&&
 -\sum_{n,m}  \  m_{{v}^n}^2 a_{{v}^n} \psi_{2m-1}
\int d^4y\
 \k K  \widehat{\mathbb{F}}_{Z \n}(y,Z) \Delta^{\n \m}_{mn}(y-x) \Big|_{Z=B}
 \ .
  \label{Bcurrent2}
\end{eqnarray}
The massive vector meson propagator in Lorentz gauge is
\begin{eqnarray}
\Delta_{\m\n}^{mn}(x) = \int \frac{d^4 p}{(2\pi)^4} e^{-ipx}
\left[ \frac{- g_{\m\n} - p_\m p_\n / m_{v^n}^2}{p^2 + m_{v^n}^2}
\d^{mn}\right] \ ,
\end{eqnarray}
The first contribution in (\ref{Bcurrent2}) is the direct coupling between
the core instanton and the $U(1)_V$ source as displayed in Fig.~2a.
The second contribution sums up the omega, omega', ... contributions
as displayed in Fig.~2b.  We note that the direct or core coupling
drops by the exact sum rule
\begin{eqnarray}
\sum_{n=1}^{\infty} \a_{v^n}\psi_{2n-1} = 1 \ , \label{sumrule1}
\end{eqnarray}
following from closure in curved space
\begin{eqnarray}
\d (Z-Z') = \sum_{n=1}^{\infty}\ \k \psi_{2n-1}(Z)
\psi_{2n-1}(Z')K^{-1/3}(Z') \ .
\label{SUM}
\end{eqnarray}
after integrating over the tower of omega meson trajectory. 
Vector Meson Dominance (VMD) is exact
in holography. A similar argument holds for the pion electromagnetic 
form factor in~\cite{SAKAI}. The results presented in this section were derived in~\cite{KZ}
using the cheshire cat descriptive. They were independently arrived
at in~\cite{HASHI} using the strong coupling source quantization.
They also support, the bottom up effective approach 
described in~\cite{RHO} using the heavy nucleon expansion.

For many years Gerry Brown has been advocating the 50/50 scenario for the baryon
form factor using both phenomenology and his democratic principle. In many ways, 
the present unwinding of the baryon current in holography supports that. Indeed, if
we were to truncate the resonance contributions to the lowests, say the omega, then 
the core contribution is non-zero. In this case, the deleniation of the Cheshire cat
smile is no longer arbitrary. A specific position of the smile, will garentee optimal 
rearrangement between the truncated cloud and the core contributions, thereby
vindicating Gerry's 50/50 scenario.

\begin{figure}[]
  \begin{center}
    \includegraphics[width=11cm]{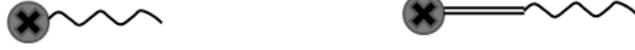}
  \caption{Gerry Brown 50/50 scenario:  Direct (left)  + VMD (right). See text.}
  \label{Fig:FF}
  \end{center}
\end{figure}

\section{2-Skyrmions from 2-Instantons}

\indent The procedure from the 1-nucleon sector to the 2-nucleon sector
in holography relies on substituting the 1-instanton by the 2-instanton
configuration.  The latter is encoded in the ADHM data $\Delta$ which is
a $(1+k)\times k$ matrix~\cite{ADHM,baal}

\begin{eqnarray}
\D = \begin{pmatrix}
       \l_1 & \l_2 \\
       D-x & u \\
       u & -D-x \\
     \end{pmatrix} \ , \qquad
     \D^{\dagger} \equiv \begin{pmatrix}
                 \l_1^\dg & (D-x)^\dg  & u^\dg \\
                  \l_2^\dg & u^\dg & (-D-x)^\dg \\
               \end{pmatrix} \ , \label{Ddagger}
\end{eqnarray}
where the coordinates $x_M$ are defined as $x=x_M \s^M$, and the moduli parameters are
encoded in the free parameters $\l_1, \l_2, D$:
$|{\l_{i}}|  \equiv  \rho_{i} $ are the size parameters,
$\l_1^{\dg}\l_2/(\rho_1\rho_2) \in SU(2)$ is the relative gauge orientation,
and $\pm D$ is the location of the constituents. $u$ is a parameter fixed by the
ADHM constraint $\Delta^\dagger\Delta=f^{-1}\otimes {\bf 1}$,

\begin{eqnarray}
f^{-1}  = \begin{pmatrix}
                   \rho_1^2 + (x_M  -  D_M)^2 + \frac{\rho_1^2\rho_2^2-
(\l_{1} \cdot \l_{2})^2}{4D_M^2} &  \l_{1} \cdot \l_{2} + 2x \cdot u \\
                   \l_{1} \cdot \l_{2} + 2x \cdot u & \rho_2^2 + (x_M  +  D_M)^2
+ \frac{\rho_1^2\rho_2^2- (\l_{1} \cdot \l_{2})^2}{4D_M^2} \\
                 \end{pmatrix}  \ ,
\end{eqnarray}
where the notation $ q \cdot p $ for two quaternions $q$ and $p$ is used, 
\begin{eqnarray}
q \cdot p \equiv \sum_M q_{M} p_{M} \ .
\end{eqnarray}
$\rho_{i} = \sqrt{ \l_{i} \cdot  \l_{i} } $ are the size parameters,
$\pm D_M$ the relative positions of the instantons, and
\begin{eqnarray}
2 x \cdot u = \frac{1}{D\cdot D} \left[ (\l_2 \cdot D)(\l_1 \cdot x)
- (\l_1 \cdot D)(\l_2 \cdot x) - \e^{MNPQ}(\l_2)_M (\l_1)_N D_P x_Q  \right] \ .
\end{eqnarray}
%We made use of the identity
%
%\begin{eqnarray}
%&& \s^P \bar{\s}^{MN} = \d^{PM}\s^{N} - \d^{PN}\s^{M} - \e^{PMNQ}\s^{Q} \ , \nn \\
%&& \bar{\s}^{MN} \equiv \half(\bar{\s}^M \s^N - \bar{\s}^N \s^M) \ , \quad \e^{1234} = 1 \ .
%\end{eqnarray}

%
\begin{figure}[]
  \begin{center}
    \includegraphics[width=11cm]{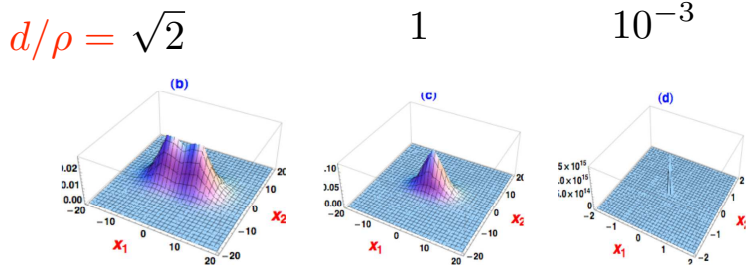}
  \caption{2-Skyrmions from 2-Instantons: Defensive}
  \label{Fig:FFD}
  \end{center}
\end{figure}

Without loss of generality, we may choose the moduli parameters to be
\begin{eqnarray}
\l_1 = \rho_1\left(0,0,0,1 \right)\ , \quad  \l_2 = \rho_2\left(\wh{\theta}_a \sin\!\abs{\,\theta} , \cos\!\abs{\,\theta} \right) \ ,
\quad D = \left( \frac{d}{2},0,0,0 \right)  \ ,  \label{parameter}
\end{eqnarray}
with $a=1,2,3$, $\abs{\,\theta} \equiv \sqrt{(\theta_1)^2
+ (\theta_2)^2 +(\theta_3)^2}$ and $\wh{\theta}_a \equiv \frac{\theta_a}{\abs{\,\theta}} $.
The spatial $x^1$ axis is chosen as the separation axis of two instantons at large distance
$d$. The flavor orientation angles ($\theta_a$) are relative to the $\l_1$ orientation.
We assign an $SU(2)$ matrix $U$  to the relative angle orientations in flavor space
\begin{eqnarray}
U \equiv \frac{\l_1^\dagger \l_2}{\rho_1 \rho_2} = e^{i\theta_a\tau^a} \in SU(2) \ .
\end{eqnarray}
which is associated with the orthogonal $SO(3)$ rotation matrix $R$ as
\begin{eqnarray}
R_{ab} &=& \half \Tr \left( \tau_a  U \tau_bU^\dagger \right) \nn \\
       &=& \d_{ab}\cos 2\!\abs{\,\theta} + 2\wh{\theta}_a \wh{\theta}_b \sin^2\!\abs{\,\theta}
+ \e_{abc} \wh{\theta}_c \sin2\!\abs{\,\theta}  \ . \label{SO3}
\end{eqnarray}

For instance $R_{ab}$ reads
\begin{eqnarray}
\begin{pmatrix}
  \cos 2\theta_3 & \sin 2\theta_3 & 0 \\
  -\sin 2\theta_3 & \cos 2\theta_3& 0 \\
  0 & 0 & 1 \\
\end{pmatrix} \ , \qquad
\begin{pmatrix}
  1 & 0 & 0 \\
  0 & \cos 2\theta_1 & \sin 2\theta_1  \\
  0 & -\sin 2\theta_1 & \cos 2\theta_1  \\
\end{pmatrix} \ ,
\end{eqnarray}
for $\theta_1 = \theta_2 = 0$ and $\theta_2 = \theta_3 = 0$ respectively.
Note the double covering in going from SU(2) to SO(3).

\begin{figure}[]
  \begin{center}
    \includegraphics[width=11cm]{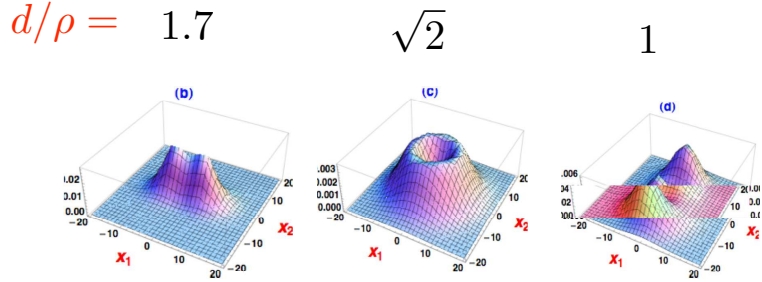}
  \caption{2-Skyrmions from 2-Instantons: Combed}
  \label{Fig:FFC}
  \end{center}
\end{figure}

In Fig.~\ref{Fig:FFD} we show the behavior of ${\rm Tr}(F^2_{\mu\nu})$
for two parallell or defensive Skyrmions with $|\theta|=0$ and $z=x_3=0$ 
and equal size cores  $\rho=9.64$ for several separations $d/\rho$. We
recall that the physical core size in units of  the KK-scale $M_K$ is
$\rho M_{K}\equiv \rho M_K/\sqrt{\lambda}$. In Fig.~\ref{Fig:FFC} we show
the behavior for two antiparallel or combed Skyrmions
with $\theta_1=\theta_2=0$ and  $\theta_3=\frac{\pi}{2}$
or $\abs{\,\theta} = {\pi}/{2}$. This is a $\pi$ rotation
along $x_3$ in the SO(3) notation (\ref{SO3}). 
For large separation two lumps form along the $x^1$ axis.  For smaller separation
the two lumps are seen to form in the orthogonal or $x_2$ direction. In between a hollow
baryon 2 configuration is seen which is the precursor of the donut seen in the baryon
number 2 sector of the Skyrme model~\cite{DONUT}. The concept of $\wt{d}$ as a separation
at small separations is no longer physical given the separation taking place in the
transverse direction. What is physical is the dual distance $u$ in the transverse plane.

\begin{figure}[]
  \begin{center}
    \includegraphics[width=11cm]{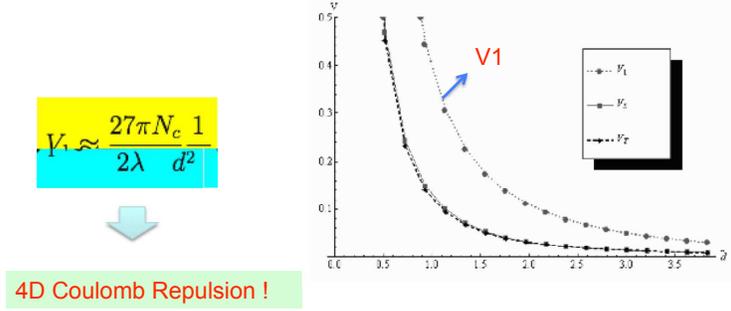}
  \caption{The central potential: Core contribution.}
  \label{Fig:CORE}
  \end{center}
\end{figure}

At large separation, the nucleon-nucleon core interaction
can be readily extracted from the Skyrmion-Skyrmion core interaction
as it is linear in the $SO(3)$ rotation $R$. Indeed, the NN-potential can be
decomposed as~\cite{KIMZAHED}.  

\begin{eqnarray}
&& V_{NN} = V^+_1 + \vec{\t}_1 \cdot \vec{\t}_2 \, V_1^-
+\vec{\s_1}\cdot\vec{\s_2} \left(V_S^+ + \vec{\t}_1\cdot\vec{\t}_2 \, V_S^- \right) \\
&& \qquad \quad +
\left(3 (\vec{\s}_1 \cdot \wh{d} ) (\vec{\s}_2 \cdot \wh{d}) - \vec{\s}_1 \cdot \vec{\s}_2 \right)
 \left(V_T^+ + \vec{\t}_1\cdot\vec{\t}_2 \, V_T^- \right)
\end{eqnarray}
with the core contribution
\begin{eqnarray}
V_{1,\mathrm{core}}^+ \approx \frac{27\pi N_c}{2\l}\frac 1{d^2} \ ,
\label{REPULSION}
\end{eqnarray}
as shown in Fig.~\ref{Fig:CORE}. This repulsion is Coulomb-like in 5-dimensions,
a hallmark of holography. The cloud contributions are meson-mediated. To lowest
order they read~\cite{KIMZAHED}

\begin{eqnarray}
&&V_{1,\wh{V}}^+   \approx   \sum_n G^2_{1\wh{V},2n-1}
 \frac{e^{-m_{2n-1}d}}{4\pi\, d}   \ ,  \qquad  G_{1\wh{V},2n-1} \equiv \frac{N_c}{2} \psi_{2n-1} \ \sim \ \sqrt{\frac{N_c}{\l}} \ ,   \\
%&&V_{1,V}^{-} \approx\sum_n G_{1V,2n-1}^2
% \frac{e^{-m_{2n-1}d}}{4\pi\,d}   \ , \, \qquad
%G_{1V,2n-1}\equiv \frac{\psi_{2n-1}}{2}  \ \ \quad  \sim \ \frac{1}{\sqrt{N_c\l}} \ ,  \\
&& V_{S,A}^- \approx  \sum_n  G_{SA,2n}^2 \frac{e^{- m_{2n}} d}{4\pi d} \ , \qquad \quad \ \
G_{SA,2n} \equiv - \frac{g_A\psi_{2n}}{\sqrt{6} \psi_0}\quad \  \sim \  \sqrt{\frac{N_c}{\l}}\ ,  \\
&& V_{T,A}^- \approx  \sum_n  G_{TA,2n}^2 \frac{e^{- m_{2n}} d}{4\pi d}  \ , \qquad \quad \ \
G_{TA,2n} \equiv  \frac{g_A\psi_{2n}}{\sqrt{12} \psi_0}\quad \ \  \sim \  \sqrt{\frac{N_c}{\l}}  \ , \\
&& V_{T,\Pi}^- \approx \frac{1}{16\pi} \left(\frac{g_A}{f_\pi}\right)^2 \inv{d^3}\ \ \sim \  \frac{N_c}{\l} \ .
\end{eqnarray}
To order $N_c/\l$ we note $V_1^-=V_S^+=V_T^+=0$.

The present description of the 1- and 2-baryon configurations resemble in many ways
the 1- and 2-nucleon structure emerging from strong coupling lattice QCD at finite
density~\cite{FORCRAN}.

\section{Dyonic Salt}

At finite density, baryons crystalize at large $N_c$ irrespective
of the 't Hooft coupling $\lambda$ since the Coulomb-like ratio is $\Gamma\approx N_c^2\gg 1$. 
The  crystal translates to a crystal of instantons in $T^3\times R_Z$ with $R_Z$ the holographic
direction.  Periodic directions are accompanied by twists or holonomies.  Twisted instantons or
instantons with holonomies are known to topologically split, if the twist or the holonomy are
strong enough. An example is the KvLL instanton in $T^1\times R^3~$\cite{baal,leeyi,lee}
which is found to split into dyons. 

\begin{figure}[]
  \begin{center}
    \includegraphics[width=11cm]{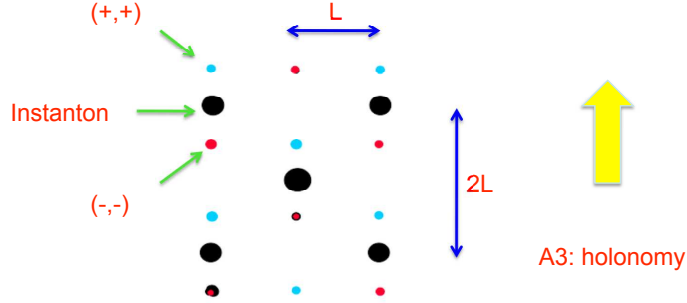}
  \caption{Instantons (black) Splitting into oppositely charged Dyons $(e,g)=(\pm ,\pm$).}
  \label{Fig:DYONS}
  \end{center}
\end{figure}

Could such a splitting take place for an instanton  arrangement in $T^3\times R$? In~\cite{SINZAHED} 
we suggested that it does, provided that the flavor gauge-symmetry is Higgsed, say the longitudinal "rho"

\be
\left<{\bf A}^3_{3}\right>=\frac{2\pi}{2L}\,v\,T^3
\label{HOLO}
\ee
develops a non-vanishing expectation value. If that is the case, and proceeding by analogy with
the KvLL instanton at finite temperature~\cite{baal,leeyi,lee,DIAKONOV}, the periodic instanton array
in space splits into a dyonic array. In Fig.~\ref{Fig:DYONS} we show how an initial crystal
arrangement of fcc dyons split into a crystal arrangement of bcc dyons under the action of the
flavor holonomy~\cite{HOLO}. The dyons are oppositely charged $e=g=\pm 1$ in units of $T^3$
along $x^3$. The dyon masses are $M_+=MB_+=Mv$ and $M_-=MB_-=M(1-v)$ -- where $v$ 
is the Higgs vev -- with $B_\pm$ their
topological charges respectively~\cite{DIAKONOV}.
We recall that $B_++B_-=v+(1-v)=1$ is the instanton
number. Here $2L$ is the cell size of the
initial fcc instanton arrangement.

\begin{figure}[]
  \begin{center}
    \includegraphics[width=11cm]{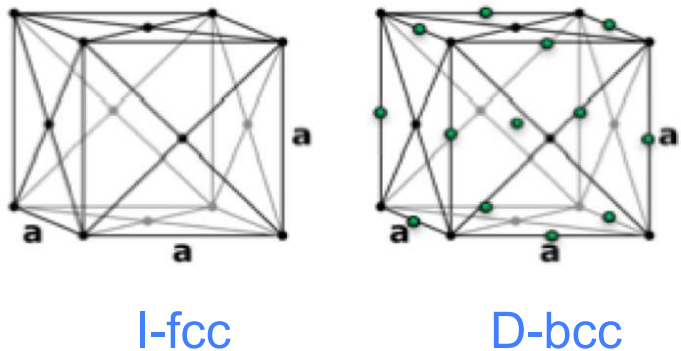}
  \caption{(a) Instantons in fcc; (b) Dyons in bcc.}
  \label{Fig:SPLIT}
  \end{center}
\end{figure}

To order $N_c\lambda\approx \kappa$
the instantons and dyons are BPS with an arbitrary value of the vev $0\leq v<1$.  The dyonic
crystal is salt-like with intertwined lattices of topological charges $v$ and $(1-v)$ at the
vertices.  In Fig.~\ref{Fig:SPLIT} we display  the fcc instanton crystal (left) as it splits to a bcc
crystal of dyons under the action of the spatial holonomy along $x^3$.

The instantons cease to be BPS at next to next to leading order (NNLO). Indeed, at NNLO
the core instantons repel at short distances as in (\ref{REPULSION}).  In general, the exact
many-dyon interaction for all ranges is involved~\cite{leeyi}.
Fortunately, for our dyonic crystal the details of the dyonic interactions are not important in
the non-BPS regime. Indeed, once the instantons split into $e=g=\pm 1$ dyons as in
Fig.~\ref{Fig:DYONS}, the Coulomb nature of the underlying charges will cause them to
arrange in a salt-like configuration to maximally screen the $+$ and $-$ charges, and therefore
balance the Coulomb forces. The Coulomb balance leaves the vev $v$ arbitrary. However,
the topological repulsion (\ref{REPULSION}) wil maintain the Coulomb balance if $v=(1-v)$.
Thus $v=1/2$ resulting into dyons of equal masses $M/2$ and equal charges $e=g=\pm 1$.
This assumption is supported by dynamical calculations using colored
instantons in SYM~\cite{DAVIS} and colored and thermal instantons in QCD in~\cite{ARIEL,MAXIM}.

The split dyon crystal arrangement is again salt-like with a unit cell $2L$ . This is a bcc crystal of
half-instantons or dyons per cubic cell $L$. The instanton or baryon density is

\be
n_B=\frac{1/2}{L^3}=\frac{4}{(2L)^3}
\label{DENS}
\ee
which is commensurate with the initial density of fcc instantons, namely a cell unit of $(2L)$
with 4 instantons. Hence our initial choice of the fcc configuration for the instantons at low
density. (\ref{DENS}) reflects on the half-instanton symmetry of the bcc dyonic salt,
which is dual to the half-skyrmion symmetry on the boundary. The density at which the
splitting takes place can be estimated using the KvLL instanton. Indeed, in the latter
dyon separation $R_{+,-}$ is~\cite{baal}

\be
R_{+-}=2\pi\,\frac{\rho^2}{2L}
\label{RPM}
\ee
with $\rho$ the KvLL instanton size with zero holonomy. Using (\ref{RPM}) for our fcc
crystal of cell size $(2L)^3$ and setting $R_{+-}=L$ at the transition to the bcc, yields
$L=\sqrt{\pi}\rho$ or a critical density $n=1/2/(\sqrt{\pi}\rho)^3$.  In hQCD, 
$\rho\approx \sqrt{2/5}\,$ fm and $L\approx 1$ fm~\cite{SINZAHED}. 
So the fcc to bcc transition takes place at $n_B\approx 1/2\,{\rm fm}^{-3}$ or 
3 times nuclear matter density.

The energy density for which the fcc to bcc transition occurs can be estimated using 
only the Coulomb crystal (ignoring the core repulsion and the meson exchange attraction).
The energy per baryon $E/N=M-\Delta$ is then~\cite{SINZAHED}

\be
\Delta=(e^2+g^2)\,(T_3)^2\,\frac{\pi}L\,M_D
%(e^2+g^2)(T_3)^2\,\frac{2\,{\rm tan}^{-1}(\rho/L)}{L}\,M_D
\label{BIN}
\ee
in the limit $\rho/L\ll 1$.  $\Delta$ is the energy to bring a dyon in a bcc configuration
and $M_D\approx 1.75$ 
is Madelung constant for salt~\cite{madelung}.  For $L\approx 1{\rm fm}$,  $\Delta\approx 
350{\rm MeV}$. This estimate is on the high side since the Madelung constant for
our 4-dimensional salt is smaller than that for the 3-dimensional salt, and 
$\rho/L\approx 1/\sqrt{\pi}$.

\begin{figure}[]
  \begin{center}
    \includegraphics[width=11cm]{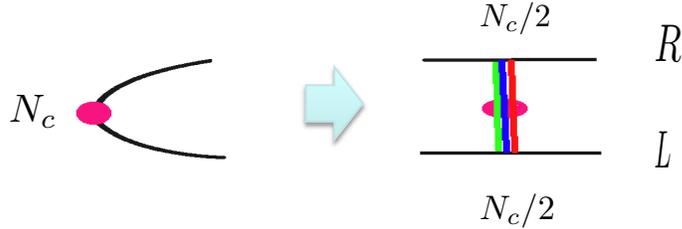}
  \caption{Geometrical reorganization of Witten's vertex from fcc (left) to (bcc) right.}
  \label{Fig:VERTEX}
  \end{center}
\end{figure}

Does the dyonic salt configuration in bulk correspond to a chirally restored phase at high density?
In~\cite{SINZAHED} we have suggested that the Witten vertex maybe re-organized from the fcc
to bcc configuration as shown in Fig.~\ref{Fig:VERTEX}.   In bulk, the probe flavor branes
$D8+\overline{D8}$ (left) split into separate $D8$ and $\overline{D8}$, each of which supporting its 
own crystal of $1/2$-instantons or dyons ($L,R$ crystals).  As a result, the right and left Wilson lines decouple
\be
U^R_{1/2}(x)= P\exp\left(i\int_0^{+\infty} \,{\bf A}_Z(x,Z)\,dZ\right), \quad\quad
U^L_{1/2}(x)=P\exp \left(i\int^0_{-\infty}\, {\bf A}_Z(x,Z)\,dZ\right) \\
\ee
The $L,R$ crystalline structures are commensurate
with the $e=g=\pm 1$ dyonic structures as both are interchangeable by parity.

\section{Dyonic Liquid}

\begin{figure}[]
  \begin{center}
    \includegraphics[width=11cm]{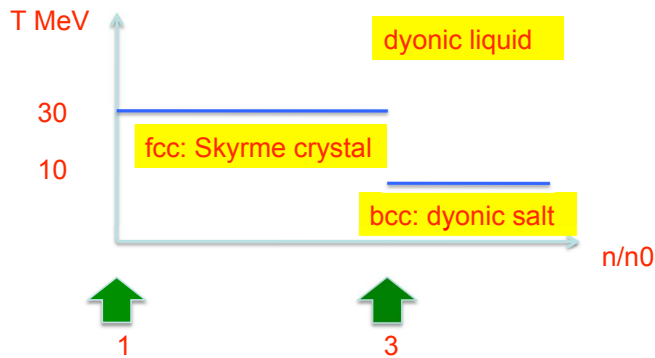}
  \caption{Phase diagram at large $N_c$ and low temperature.}
  \label{Fig:MELTING}
  \end{center}
\end{figure}

The cold dyonic crystal discussed in the preceding section can be dissociated by quanum fluctuations
and/or thermal effects, both of which are subleading in the holographic counting. This notwithstanding, 
we may qualitatively ask how much temperature will be necessary to melt the instanton or dyon crystals.
For that, Lindemann criterion tells us that when the vibration amplitude of the lattice ion reaches 
$10\%$ of the nearest neighbor distance $a_{NN}$, the classical lattice melts~\cite{LINDEMANN}

\be
\sqrt{\left<x^2\right>}\approx 10\%\,a_{NN}
\label{LIND}
\ee
The mean-square vibration can be estimated from the Einstein formulae

\be
M\omega_E^2\left<x^2\right>=k_BT
\label{EINS}
\ee
with $\omega_E\approx c_Sk_{\rm max}=c_S\pi/a_{NN}$ the Einstein frequency. The speed of sound $c_S$ in
the crystal is related to its bulk compressibility ${\bf K}$,

\be
\frac{c_S^2}{c^2}=\frac{{\bf K}}{n\,Mc^2}\approx \left(0.2-0.3\right)\frac {n_{NM}}{n}
\label{COMP}
\ee
with $n_{NM}$ again the nuclear matter density and $n$ the crystal baryon density. The last
estimate borrows the typical nuclear matter compressibility estimate 
${\bf K}/n_{NM}\approx 200-300\,{\rm MeV}$~\cite{KMODULUS}. Inserting (\ref{EINS}) and (\ref{COMP}) in
the Lindemann criterion (\ref{LIND}) yields the estimated melting temperature

\be
\frac{k_BT}{Mc^2}\approx \pi^2\,\left(10\%\right)^2\,\left(0.2-0.3\right)\,\frac{n_{NM}}{n}
\label{MELT}
\ee
The higher the crystal density, the easier to melt by this estimate. We show in 
Fig.~\ref{Fig:MELTING} a typical structure of the phase diagram at low temperature
and large $N_c$. The fcc crystals melts to a Skyrmion liquid at about 30 MeV, while the dyonic
salt melts into a dyonic liquid at about 10 MeV. The low nature of these melting temperatures
is an early indication that these crystals are easily dissociated by quantum vibrations due to 
kinetic energy for instance.

\section{Conclusions}

In hQCD baryons emerge as holograms of instantons in a curved
five-dimensional space. The baryons embody the essentials of the 
Cheshire cat principle, with the Cheshire cat found "hiding" in the 
holographic direction. A baryon with a small core and a rich cloud
of vector mesons is unravelled. Exact vector dominance emerges
when the entire Regge trajectory of the vectors is kept in the cloud.

The two-baryon configuration is naturally obtained from the two-instanton
configuration in bulk using the ADHM construction. Many features of the
Skyrmion-Skyrmion interactions are this way recovered. The adventage
of hQCD over the Skyrme model is the fact that the boundary action for
the baryons is entirely fixed by the gauge-gravity in bulk using the
large $N_c$ and large $\l$ bookeeping. 

Cold dense nucleonic systems crystallize at large $N_c$. The crystals are
found to be an arrangement of instantons in the fcc configuration at low
densities, and a bcc arrangement of dyons at higher densities. The transition
occurs at about 3 times nuclear matter densities in hQCD. Quantum kinetic
effects or thermal effects are likely to turn each of the crystal into a strongly
interacting liquid of instantons for the former and dyons for the latter.

\section{Acknowledgments}
I thank K.Y. Kim, M. Rho and S.J. Sin for the many discussions during our
long term collaboration on the issues discussed here.
This work was supported in part by US-DOE grants 
DE-FG02-88ER40388 and DE-FG03-97ER4014.

\end{document}